\documentclass[preprint,nofootinbib]{revtex4}
\usepackage{graphicx}
\usepackage{subfigure}
\usepackage{epsfig}
\usepackage{amssymb}
\begin{document}

\newcommand{\lP}{l_{\mathrm P}}

\newcommand{\md}{{\mathrm{d}}}
\newcommand{\tr}{\mbox{tr}}

\newcommand*{\R}{{\mathbb R}}
\newcommand*{\N}{{\mathbb N}}
\newcommand*{\Z}{{\mathbb Z}}
\newcommand*{\Q}{{\mathbb Q}}
\newcommand*{\C}{{\mathbb C}}

\newcommand{\ket}[1]{| #1 \rangle}
\newcommand{\bra}[1]{\left\langle #1 \right|}
\newcommand{\braket}[2]{\langle #1 \,|\, #2 \rangle }
\newcommand{\braketfull}[3]{\langle #1 \,|\, #2 \,|\, #3 \rangle }
\newcommand{\dualbra}[1]{( #1 |}
\newcommand{\dualket}[1]{| #1 )}
\newcommand{\dualbradualket}[2]{( #1 \,|\, #2 ) }

\newcommand{\dualbraket}[2]{( #1 \, | \, #2 \rangle }

\title{On the Physical Hilbert Space of Loop Quantum Cosmology}
\author{Karim Noui}
\email{karim@lmpt.univ-tours.fr}
\address{Laboratoire de Mathematiques
et Physique Theorique
Parc de Grandmont 37200 Tours France}
\author{Alejandro Perez}
\email{perez@gravity.psu.edu}
\author{Kevin Vandersloot}
\email{kfvander@gravity.psu.edu}
\address{Institute for Gravitational Physics and Geometry, The Pennsylvania
State University, 104 Davey Lab, University Park, PA 16802, USA}

\vspace{.5cm}
\begin{abstract}
In this paper we present a model of Riemannian loop quantum
cosmology with a self-adjoint quantum scalar constraint. The
physical Hilbert space is constructed using refined algebraic
quantization. When matter is included in the form of a
cosmological constant, the model is exactly solvable and we show
explicitly that the physical Hilbert space is separable consisting
of a single physical state. We extend the model to the Lorentzian
sector and discuss important implications for
standard loop quantum cosmology.
\end{abstract}

\maketitle

\section{Introduction}

Symmetry reduced models of General Relativity have been
extensively studied and applied in the framework of cosmology. In
cosmological models the infinite degrees of freedom of the full
theory are reduced to a finite number describing the large scale
structure of the universe.  Current observations seem to be best
described by (spatially flat) homogeneous and isotropic
Friedman-Robertson-Walker space-times. For that reason, the
assumption of homogeneity and isotropy is by now an accepted basic
ingredient in the so-called {\em standard model} of cosmology.

Under quite general conditions, general relativity predicts the
development of space-time singularities. One prime example of such
singular behavior is the presence of the initial Big-Bang
singularity in cosmological models. Energy densities as well as
space time curvature diverge approaching the initial Big-Bang
singularity, and any description based on classical space time
notions becomes inapplicable. General Relativity is no longer
valid and has to be replaced by a quantum theory of gravity.

Although no complete quantum theory of gravity is yet available,
loop quantum gravity appears as an promising
candidate\cite{book,bookt,ash10,ap}. Loop quantum gravity is a
background independent non perturbative approach to the
quantization of general relativity.
The main prediction of the theory is the discreteness of geometry
at the fundamental level: operators corresponding to geometric
quantities such as the area and volume have discrete spectra
\cite{lee1,c3}. Geometry is quantized and at small scales the
smooth notion of space and time simply can no longer be applied.
One of the main clear-cut applications of the theory, where the
discovery of discreteness plays an important role, is the
description of the fundamental degrees of freedom of black hole
(isolated)
horizons\cite{c5,Krasnov:1996wc,Ashtekar:1997yu,Ashtekar:2000eq,Ashtekar:1999wa,Meissner:2004ju,Domagala:2004jt}.
These fundamental excitations give rise to the correct value of
black hole entropy that is expected from semi-classical
considerations.

The application of loop quantum gravity to cosmological models is
known as loop quantum cosmology (LQC) \cite{Bojowald:2004ax} (See
also \cite{Gam} for an alternative approach). The
simplification arising from the symmetry reduction makes it possible
to address physical questions that still remain open in the full
theory. For instance LQC provides a novel paradigm for the
understanding of the fate of the Big-Bang singularity in quantum
gravity.  The important physical question of whether classical
singularities of general relativity are not present in the quantum
theory is answered in the affirmative in
LQC\cite{Bojowald:2001xe}. The basic mechanisms leading to this
striking result can be traced back to the fundamental discreteness
of geometric operators which drastically changes the quantum
evolution equation in the deep Planckian regime. In
addition, consistency conditions,known as the dynamical initial
condition, seem to severely restrict the
freedom in the choice of `initial' values for the evolution and it
is argued that these conditions may single out a unique wave
function of the universe\cite{Bojowald:2001xa}. It has also been
shown that loop quantum cosmology leads to modifications of the
classical Friedman equation. Recent work has shown that these
effects can provide the proper initial conditions for chaotic
inflation and may lead to measurable effects on the CMB
\cite{Tsujikawa:2003vr}.

In spite of these successes, a complete characterization of the
physical Hilbert space of loop quantum cosmology remains an open
issue\footnote{This remains an open issue in the full
theory of LQG. The master constraint approach \cite{Thiemann:2003zv} and
the spin foam formulation \cite{Perez:2003vx} are current proposals
to solve this problem.}.
The definition of the physical inner product is the missing
piece of the puzzle in LQC. Without the physical inner product,
there is no notion of a probability measure which is an essential
ingredient for physical prediction in the quantum theory. Any
interpretation of the wave function as a probability amplitude is
futile without the probability measure provided by the physical
inner product.

In addition the physical inner product provides a means
for eliminating pathological solutions to the
constraints. For instance it can be shown that the
space of solutions of LQC is generally non separable, i.e., an
uncountable infinity of solutions to the quantum constraint
equation exist.
Such a huge physical Hilbert space is certainly not expected for a
system with finite number of degrees of freedom. Some of
these solutions exhibit peculiar properties such as wild
oscillatory or diverging behavior in the regime
where one would expect agreement with standard Wheeler-DeWitt
quantum cosmology. In the absence of the
physical inner product, so far spurious solutions are eliminated
using the dynamical initial condition coupled with
heuristic arguments based on semi-classicality requirements.
Although these requirements are physically motivated,
unphysical solutions can be easily identified if a notion of
physical inner product is provided. More precisely, on the basis
of the notion of physical probability physical states are defined
by equivalence classes of solutions up to zero norm states. It is
hoped that the ill-behaved solutions mentioned above will be
factored out by this means once a suitable notion of physical
inner product is provided.

There exists a simple method known as group averaging
to provide a definition of the
physical inner product in generally covariant systems when the
constraints generate a group action
\cite{Marolf:2000iq}. In this paper we review and use the
technique called {\em refined algebraic quantization} \cite{ash5}
as a recipe to construct the physical Hilbert space. The physical
inner product is constructed  in
a simple manner if one can define a unitary operator generating
`translations' along the gauge orbits generated by the quantum
constraints in a suitable kinematical Hilbert space.
 This technique cannot be applied to the standard
formulations of loop quantum cosmology because the quantum
Hamiltonian constraint is not a self adjoint operator (in the
kinematical Hilbert space) and thus it does not lead to any well
defined unitary `evolution' when exponentiated.

In this paper we present a model of loop quantum gravity that
arises from a symmetry reduction of the self dual Plebanski
action. In this formulation the symmetry reduction leads to a very
simple Hamiltonian constraint that can be quantized explicitly in
the framework of loop quantum cosmology. The model is defined in
the Riemannian sector which constitutes a limitation when
considering physical predictions. However, we will present a
method for transforming to the Lorentzian sector and then argue
that the main results of the model are manifested in the
Lorentzian sector. The model shares the same kinematical Hilbert
space with standard loop quantum cosmology, but differs in that
the resulting Hamiltonian constraint is self-adjoint and thus the
group averaging technique can be used to construct the physical
inner product. When matter is incorporated in the form of a
cosmological constant, we can solve the model exactly both for the
physical wave functions and the physical inner product. The
physical inner product reduces the infinite set of solutions of
the quantum Hamiltonian constraint equation to one physically
normalizable solution. This constitutes an interesting example
where the very large (non separable) kinematical Hilbert space is
reduced to a (trivially) separable physical Hilbert space. 

We end with a discussion of the relevance of the model to
standard LQC. We show that despite the simpler constraint
the model shares the same qualitative features of standard
LQC with one important difference. 
We will then discuss the implications that can
be derived about LQC in general.

\section{Classical Symmetry Reduction}
Our starting point is the classical self dual Riemannian
gravitational action which, as noted by Plebanski\cite{pleb}, can
be written as a constrained SU(2) BF theory
\begin{equation}\label{full_BF_action}
    S[A,\Sigma, \psi] = \int  \Sigma_i \wedge F^i(A)  -
    \psi_{ij} \left[ \Sigma^i \wedge \Sigma^j -
        \frac{1}{3} \delta^{ij} \Sigma^k \wedge \Sigma_k \right]
    + \Lambda \, \Sigma_i \wedge \Sigma_i
\end{equation}
where $\Sigma=\Sigma^i_{\mu \nu} dx^{\mu} dx^{\nu} \tau_i$ is
an SU(2)
Lie algebra valued two form, $A=A^i_{\mu} dx^{\mu} \tau_i$ is an 
SU(2) Lie algebra
valued connection, $\tau_i$ are the generators of the Lie algebra
of SU(2), and $\Lambda$ is a cosmological constant.
In this paper we set the Planck length $l_p=\sqrt{8\pi G \hbar}$ equal to one
and restrict the model to flat k=0 cosmologies.
The tensor $\psi_{ij}$ is symmetric and acts
as a Lagrange multiplier enforcing the constraint
$ \Sigma^i \wedge \Sigma^j -
\frac{1}{3} \delta^{ij} \Sigma^k \wedge \Sigma_k = 0$.
Once this constraint is solved, the action becomes equivalent to
the self-dual action of general relativity.

In order to reduce the action (\ref{full_BF_action}) to spatial
homogeneity and isotropy, we write the action explicitly in terms
of coordinates separating space and time as
\begin{eqnarray}\label{split_BF_action}
    S_{GR}[A,\tilde{E},B] = \int\!\! dt \int\!\! & d^3\!x &
    \dot A^i_a \tilde{E}^a_i
        +A^i_0 D_a \tilde{E}^a_i
        +\epsilon^{abc} B^i_a F^i_{bc}  \nonumber \\
    &-& \psi_{ij}  \left[\frac{1}{2}  B^i_a \tilde{E}^{aj} +
        \frac{1}{2} B^j_a \tilde{E}^{ai} -
    \frac{4}{3} \delta^{ij} B^k_a \tilde{E}^a_k \right]
        \nonumber \\
    &+& \Lambda\, B^i_a \tilde{E}^a_i
\end{eqnarray}
where we have introduced the definitions
$\tilde{E}^{ai} \equiv 2 \, \epsilon^{abc} \Sigma^i_{bc}$,
$B^i_a \equiv 2 \, \Sigma^i_{0a}$, and $\epsilon^{abc}=\epsilon^{0abc}$.
Spatially homogeneous and isotropic connections can be
characterized as\cite{Bojowald:1999eh}
\begin{equation}\label{isotropic_connection}
    A^i_a = A \Lambda^i_I \omega^I_a
\end{equation}
where the parameter $A$ encodes the gauge invariant part of the
connection, $\Lambda^i_I$ is an SO(3) matrix, and
$\omega^I=\omega^I_a dx^a\, (I=1,2,3)$ are left invariant one forms with
respect to the translational symmetry associated with homogeneity.
Since we are restricting ourselves to spatially flat cosmologies,
the translation symmetry group is isomorphic to $\mathbb{R}^3$.
Due to isotropy the time component of the connection $A_0^i$
vanishes identically
\footnote{According to the general analysis of symmetric
connections for arbitrary symmetry groups \cite{Bojowald:1999eh}
the time component of the connection is a connection
on a symmetry reduced principal fiber bundle. This
fiber bundle has as its gauge group the centralizer
$Z_{\lambda} := Z_G(\lambda(F))$ where $G$ is the
gauge group of the theory, $F$ is the isotropy
group, and $\lambda$ is a homomorphism from
$F$ to $G$. For our model the gauge group
as well as the isotropy group
is SU(2). The homomorphism
$\lambda$ is thus the identity map and
the centralizer $Z_{\lambda}$ only contains the identity. The
time component of the connection thus vanishes.
}
. The matrix $\Lambda^i_I$ is pure gauge and
satisfies
\begin{eqnarray}
    \Lambda^i_I \Lambda^I_j &=& \delta^i_j \nonumber \\
    \epsilon_{ijk} \Lambda^i_I \Lambda^j_J \Lambda^k_K &=& \epsilon_{IJK}.
\end{eqnarray}
Using left invariant vector fields
$\omega_I = \omega^a_I \partial_a$ dual to $\omega^I$ we can write the momentum canonically
conjugate to the connection as
\begin{equation}\label{isotropic_triad}
    \tilde{E}^a_i = \sqrt{q_0} E \Lambda_i^I \omega^a_I
\end{equation}
where the density weight of $\tilde{E}^a_i$ is provided by the left invariant
metric ${q_0}_{ab} = \omega^I_a \omega^J_b \delta_{IJ}$. Once again, the
gauge invariant part of the canonical momentum is given by a
single parameter $E$.
Similarly $B^i_a$ is given by
\begin{equation}\label{isotripic_B}
    B^i_a = B \Lambda^i_I \omega^I_a.
\end{equation}

Written in terms of the reduced variables the homogeneous and isotropic
gravitational action becomes
\begin{eqnarray}\label{isotropic_action}
    S_{GR}[A,E,B] &=& \int\!\! dt \int\!\!  d^3\!x
    \;3 \sqrt{q_0}  E \dot A + 6 \sqrt{q_0}  B A^2
    +3  \sqrt{q_0} \Lambda BE \nonumber \\
    &=& V_0 \int\!\! dt \; 3 E \dot A + 6 B A^2 + 3 \Lambda BE
\end{eqnarray}
where $V_0 \equiv \int \!\!  d^3\!x \sqrt{q_0}$. The constraint
term
$ \psi_{ij} \left[ B^i_a \tilde{E}^{aj} + B^j_a \tilde{E}^{ai} -
        \frac{1}{3} \delta^{ij} B^k_a \tilde{E}^a_k \right]$
in the full action (\ref{split_BF_action}) vanishes identically
for the isotropic model which can be checked by a direct
calculation. Technically $V_0$ diverges, since the
left invariant metric is constant and the integral is over a non
compact manifold. To overcome this, we restrict the integral to a finite
cell with volume $V_0$ and absorb this factor into the variables
as $E \rightarrow V_0^{2/3}E$, $A \rightarrow V_0^{1/3} A$,
and $B \rightarrow V_0^{1/3} B$, whence the action becomes
\begin{eqnarray}\label{isotropic_action2}
        S_{GR}(A,E,B) = \int\!\! dt \; 3 E \dot A + B\,( 6A^2 +3 \Lambda E) .
\end{eqnarray}
It is clear from the action (\ref{isotropic_action2}) that
$E$ is canonically conjugate to $A$ with Poisson bracket
$\{A,E\}=\frac{1}{3}$, and $B$ acts as a Lagrange multiplier enforcing
the Hamiltonian constraint $H =0$ which here for
the gravitational part is just $H=6A^2 + 3 \Lambda E$. Counting degrees of
freedom we have two dynamical variables $A$ and $E$ with one
first class constraint $H$ thus giving zero degrees of freedom
as expected for isotropic general relativity with no matter.

We now show that the action (\ref{isotropic_action2}) is equivalent
to the standard isotropic action written in terms of a single scale
factor $a$ and lapse $N$ with metric $ds^2 = N^2 dt^2 + a^2 dx^2$.
In the full BF theory, the densitized metric is given by the
Urbantke formula \cite{Urbantke:1984eb}
\begin{equation}\label{Urkantke_metric}
    \sqrt{g} g_{\mu \nu} = \frac{2}{3} \epsilon_{ijk}
    \epsilon^{\alpha \beta \gamma \delta}
    \Sigma^i_{\mu \alpha} \Sigma^j_{\beta \gamma}
    \Sigma^k_{\delta \nu}.
\end{equation}
Reducing this to isotropy gives
\begin{eqnarray}\label{istropic_metric}
    g_{00} &=& \frac{2 B^2}{E} \nonumber \\
    g_{aa} &=& \frac{E}{2}.
\end{eqnarray}
We thus have the relationships between the two sets of variables
\begin{eqnarray}
    E &=& 2 \,a^2 \nonumber \\
    B &=& N a.
\end{eqnarray}
In the full theory, the connection is a sum of the spin connection
plus the extrinsic curvature. Since we are only considering
spatially flat models the spin connection vanishes and the connection
is equal to the extrinsic curvature which implies for the isotropic model
\begin{equation}
    A = \frac{\dot{a}}{2N} \;\;.
\end{equation}
With the conventions chosen the action (\ref{isotropic_action2}) becomes
\begin{equation}
    S_{GR} = \int\!\! dt \; \frac{3}{N} \left[ a^2 \ddot{a}
            +\frac{1}{2}a \dot{a}^2
            \right] + 6 \Lambda N a^3
\end{equation}
which coincides with
the standard isotropic action of general relativity
up to a total derivative.

\section{Quantum Theory}
We wish to quantize the symmetry reduced theory in the same manner
as loop quantum cosmology by using techniques from the full theory
adapted to the symmetry. Thus, the Hamiltonian constraint is
represented as an operator on a kinematical Hilbert space using
holonomies and fluxes as the basic variables. Physical wave
functions are those annihilated by the constraint operator. In
addition refined algebraic quantization provides a means for
constructing the physical Hilbert space. In this section we start
with a formal introduction to refined algebraic quantization (or
the group averaging technique) used to calculate the physical inner
product. Next we carry out the program for the cosmological model
considered here.

\subsection{Refined Algebraic Quantization}\label{rac}
The goal of refined algebraic quantization is to find a method for
describing the physical Hilbert space, the space of wave functions
annihilated by the constraint and normalizable with a physical
inner product. Refined algebraic quantization provides a means for
determining both. We start with the formal definitions and
notation. The discussion will not serve as a general review 
of the method, but will instead be tailored to the single
constraint system at hand.

The idea of refined algebraic quantization is to build the
physical Hilbert space $\mathcal{H}_{phys}$
starting from a kinematical Hilbert space  $\mathcal{H}_{kin}$
acquired by quantizing the theory first ignoring the
constraint. The constraint is  represented as
a self-adjoint operator on this kinematical Hilbert
space. As in standard Dirac quantization, the physical states are
annihilated by the constraint operator. If the
eigenstates of the constraint operator are
not normalizable then the physical states
lie outside of the kinematical Hilbert
space. In this scenario refined algebraic
quantization provides a recipe for constructing
the physical inner product and characterizing the physical
Hilbert space.

For the case where the eigenstates of the constraint operator are
not normalizable the physical inner product is constructed
as follows. A dense subspace $\Phi \subset \mathcal{H}_{kin}$
of the kinematical Hilbert
space is chosen and solutions to the constraint equation then
lie in the topological dual $\Phi^{\star}$ of $\Phi$. We
denote the action of $\Phi^{\star}$ on $\Phi$ using
Dirac notation; namely, given an element $\bra{\psi} \in \Phi^{\star}$ and
an element $\ket{\phi} \in \Phi$ the action is denoted
$\braket{\psi}{\phi}$.
For the construction of the physical inner product, an anti-linear map
$P: \Phi \rightarrow \Phi^{\star}$ is required such that
given a state $\ket{\psi_0} \in \Phi$, $\bra{P(\psi_0)} \in \Phi^{\star}$ is
a solution to the constraint equation in the sense that
$\braketfull{P(\psi_0)}{\hat{H}}{\phi_0}=0$ for any
$\ket{\phi_0} \in \Phi$.
Thus $P$ maps onto the
kernel of the constraint operator which consists of
elements of $\Phi^{\star}$. Technically
$P$ is not a projector since $P^2$ is ill defined. 
The spaces involved satisfy the triple relation
$\Phi \subset \mathcal{H}_{kin} \subset \Phi^{\star}$.

With the map $P$ we can now define the physical inner product
and characterize the physical Hilbert space.
The physical inner product between two
states  $\bra{P( \phi_0)}, \, \bra{P(\psi_0)} \in \Phi^{\star}$ is given by
\begin{equation}\label{physical_inner_product}
    \braket{P  (\phi_0)}{P (\psi_0)}_{phys} := \;
        \braket{P(\psi_0)}{\phi_{0}} \,.
\end{equation}
where the l.h.s is just notation, and due to the anti-linearity of $P$ the order
has been reversed on the r.h.s..
A unique physical state is labeled by an equivalence class
of states in $\Phi \subset \mathcal{H}_{kin}$.
Two states $\ket{\psi_0}$ and
$\ket{\psi'_0}$ in $\Phi$ are equivalent if
\begin{equation}\label{equivstates}
    \bra{\psi_0} = \bra{\psi'_0} + \bra{x_0}
\end{equation}
for some $\ket{x_0} \in \Phi$ satisfying
\begin{equation}\label{0states}
    \braket{P( x_0)}{x_0} = 0 ,
\end{equation}
ie. when $\bra{x_0}$ has zero physical norm.
Under these conditions the physical inner product
is independent of the state $\ket{\psi_0} \in \Phi$ used
to represent the physical state in $\bra{P(\psi_0)} \in \Phi^{\star}$.
The elements of the physical Hilbert space 
can therefore be labeled by an equivalence class
of states in $\Phi$ as defined above.

So far, to construct the physical Hilbert space we have
demanded the existence of the map $P$ without providing
a method of calculating it.
We now describe the group averaging technique which provides such a method.
The map is given by
\begin{equation}\label{ga_projector}
    \bra{P(\phi_0)} = \int \! dT \bra{\phi_0} e^{-i \hat{H} T} \,.
\end{equation}
Clearly $P$ maps onto the  kernel
of the Hamiltonian constraint, that is
\begin{equation}
        \braketfull{P(\psi_0)}{\hat{H}}{\phi_0}=0
\end{equation}
for any state $\ket{\phi_0} \in \Phi$.
This can be shown by inserting a complete set of eigenstates
of the constraint operator $\hat{H} \ket{E_n} = E_n \ket{E_n}$ to
get
\begin{eqnarray}
        \braketfull{P(\psi_0)}{\hat{H}}{\phi_0} &=&
                 \int \! dT \, \braketfull{\psi_0}{ e^{-i \hat{H} T}\;
                 \hat{H}}
                {\phi_0}
                \nonumber \\
        &=& \sum_n \int \! dT \, \braket{\psi_0}{E_n} \bra{E_n}
                e^{-i \hat{H} T}  \;\hat{H}
                \ket{\phi_0} \nonumber \\
        &=& \sum_n \braket{\psi_0}{E_n} \int \! dT \,
                e^{-i E_n T}\,E_n\, \braket{E_n}{\phi_0}
                \nonumber \\
        &=& \sum_n \braket{\psi_0}{E_n}\; E_n\; \delta(E_n) \,
        \braket{E_n}{\phi_0} = 0
\end{eqnarray}
where in the case where the eigenstates are not normalizable
the sum over $n$ is replaced by an integral.

If the eigenstates are normalizable then the construction
of the physical Hilbert space is straightforward.
The map $P$ now is a bona fide operator on the kinematical
Hilbert space and behaves as a projector satisfying $P^2=P$
projecting onto
the kernel of the constraint.
The physical inner product becomes
\begin{eqnarray}
    \braket{P \, [\phi_0]}{P \, [\psi_0]}_{phys} &=&
        \braketfull{\psi_{0}}{P}{\phi_{0}} \\
    &=& \braketfull{\psi_{0}}{P^2}{\phi_{0}} 
\end{eqnarray}
thus the physical inner product is equivalent to the kinematical one
in the restriction of $\mathcal{H}_{kin}$ to the kernel of $\hat{H}$.
The definition of the equivalence relation remains the same
with a simplification arising
since $P^2=P$. 

\subsection{Kinematical Hilbert Space of Loop Quantum Cosmology}
We now turn to the cosmological model at hand.
The first step required in refined algebraic quantization
is to build the kinematical Hilbert space by
ignoring any constraints.
The kinematical Hilbert space for isotropic loop quantum 
cosmology has been constructed in \cite{Ashtekar:2003hd}.
We review the results here. 

Using ideas from
the full theory, configuration variables are constructed
from holonomies. For homogeneity and isotropy the holonomy algebra
consists of the set of almost periodic functions; namely, those that can be written
\begin{equation}
    f(A) = \sum_j f_j \; e^{i \frac{\nu_j A}{2}}
\end{equation}
where the sum contains a finite number of terms ($j=1,2,\cdots N$ and $N<\infty$),
$f_j \in \mathbb{C}$, and $\nu_j \in \mathbb{R}$. The
momentum variables consist of fluxes of the triad operator
$E^a_i$ on a 2-surface which after symmetry reduction
are trivially proportional to the parameter $E$. The kinematical Hilbert space is defined
by representing this algebra using
the Bohr compactification of the real line $\overline{R}_{Bohr}$. $\overline{R}_{Bohr}$ is
a compact Abelian group and is equipped with a Haar measure
denoted by $d\mu$ with elements being labeled
by $A\in \mathbb{R}$. The Haar measure can be explicitly
written as
\begin{equation}\label{measure}
    \int \! d\mu = \lim_{L \rightarrow \infty} \; \frac{1}{2 L}
    \int_{-L}^L dA .
\end{equation}
The kinematical Hilbert space consists of functions
which are square integrable with the Haar measure $d\mu$
which corresponds to 
the completion
of the set of almost periodic functions
in the norm of this measure.
In the notation of section \ref{rac},
$\Phi$ is the space of almost periodic
functions.
With this measure an orthonormal basis consists of states $\ket{\nu}$
given by
\begin{equation}
    \braket{A}{\nu} = e^{i A \nu /2}
\end{equation}
where the parameter
$\nu$ spans the entire real line. The basis states are normalizable
in contrast to a standard quantum mechanical representation
and satisfy the orthonormality condition
\begin{equation}
    \braket{ \nu'}{ \nu} = \delta_{\nu' \nu}
\end{equation}
which can be shown using the measure (\ref{measure}). The
basis states $\ket{\nu}$ are eigenstates of the triad
operator $\hat{E} = -i\frac{1}{3} \frac{\partial}{ \partial A}$:
\begin{equation}
    \hat{E} \; \ket{ \nu} = \frac{\nu}{6} \; \ket{\nu} .
\end{equation}
Geometrical operators can be built from the triad operator.
In particular the volume operator is given by
$\hat{V}=|\hat{E}|^{3/2}$ thus $\ket{\nu}$ represents
a quantum state with definite volume given by
$(\frac{|\nu|}{6})^{3/2}$. This provides the physical interpretation
of the label $\nu$. Finally we can write the decomposition
of the identity for both connection and triad representation as
\begin{eqnarray}\label{identity}
    1 &=& \int d\mu \;  \ket{A} \bra{A}  \nonumber \\
        1 &=& \sum^\infty_{\nu=-\infty} \ket{\nu} \bra{\nu}
\end{eqnarray}
where the sum over $\nu$ is a \emph{continuous} sum over
all values of $\nu$ on the real line. It is important
to note that the kinematical Hilbert space
here is non-separable being spanned by a non-countable
set of basis states $\ket{\nu}$.

\subsection{Quantization of the Hamiltonian Constraint}
The homogeneous and isotropic model consists of the Hamiltonian
constraint  $H_{class}=6A^2+3\Lambda E +\frac{1}{\sqrt{E}} H_{matter}$
\footnote{The factor of $1/ \sqrt{E}$ is added such
that $H_{matter}$ agrees with the standard form of the matter Hamiltonian
used in other formulations. Note
that the constraint used here does not have units of
energy hence the need for the extra factor.}.
In representing the constraint as a
self-adjoint operator on the kinematical Hilbert space
no operator corresponding to the connection $\hat{A}$ exists; instead connection
operators need to be represented in terms of the basic
variables which are holonomies along
edges. In particular the curvature term in the full action
(\ref{split_BF_action}) , which contributes the $A^2$ term after
symmetry reduction, can be represented with holonomies
around closed loops. Because of isotropy we can consider
holonomies around squares on the manifold with edge lengths
chosen to be $\nu_0 V_0^{1/3}$ for some positive parameter
$\nu_0$. The holonomy along an edge generated by the
left invariant vector field $\omega_I$ is given by
\begin{eqnarray}
    h_I &=& \exp(\nu_0 A \Lambda_I) =
     \cos(\nu_0 A/2) \;+\; 2 \Lambda_I \sin(\nu_0 A/2).
\end{eqnarray}
The $6 B A^2$ term in the action is then represented
as
\begin{equation}
    6 B A^2 \rightarrow - \frac{2}{ \nu_0^2} \; \epsilon^{IJK}
        \tr \left[  B \Lambda_I
        h_J h_K (h_J)^{-1} (h_K)^{-1} \right]=
        \frac{24 B}{\nu_0^2} \sin^2(\nu_0 A/2)\; \cos^2(\nu_0 A/2).
\end{equation}
In the limit where the closed loops are shrunk to a point by taking
$\nu_0$ to zero, the classical expression $6 B A^2$ is recovered.
In the quantum theory since there is no operator corresponding to
the connection $\hat{A}$
we cannot remove $\nu_0$ and thus
the parameter remains as a quantum ambiguity. In this paper
we will not fix it's value, but it has been argued that
it's value can be determined from the full theory of loop
quantum gravity to be equal to $\sqrt{3}/4$ \cite{Ashtekar:2003hd}.

The gravitational part of the Hamiltonian constraint
thus is given by
\begin{equation}
    \hat{H}_{GR}= \frac{24}{\nu_0^2} \sin^2(\nu_0 \hat{A}/2)
        \cos^2(\nu_0 \hat{A}/2)
        + 3 \Lambda \hat{E} \, .
\end{equation}
To determine it's action on the basis states $\ket{\nu}$ we
 first consider the action of $\sin(\nu_0 \hat{A}/2)$ and
$\cos(\nu_0 \hat{A}/2)$. Since we have $\braket{A}{\nu}
=\exp(i A \nu /2)$ we have
\begin{eqnarray}
    \sin(\nu_0 \hat{A}/2) \; \ket{\nu} &=& \frac{1}{2i} \Big( \;
    \ket{\nu \!+\! \nu_0} - \ket{\nu \!-\! \nu_0} \; \Big) \nonumber \\
    \cos(\nu_0 \hat{A}/2) \; \ket{\nu} &=& \frac{1}{2} \Big( \;
        \ket{\nu \!+\! \nu_0} + \ket{\nu \!-\! \nu_0} \;\Big)
\end{eqnarray}
and thus the action of $\hat{H}_{GR}$ is
\begin{equation}\label{constraint_action}
    \hat{H}_{GR} \, \ket{\nu} = -\frac{3}{2\nu_0^2}
    \Big( \; \ket{\nu \!+\! 4\nu_0} -2 \ket{\nu}
    + \ket{\nu \!-\! 4\nu_0} \;\Big) + \Lambda \frac{\nu}{2} \ket{\nu} \,.
\end{equation}
The action of the curvature part of the constraint operator is
thus seen to be a discrete
approximation to an operator corresponding to the second derivative
with respect to $\nu$ and in the limit of small $\nu_0$ approaches
what would be the standard quantum mechanical operator $6 \hat{A}^2 =
-24 \frac{\partial^2}{\partial \nu^2}$.

\subsection{Path Integral Representation of $P$}
Now we explicitly compute the group averaging formula (\ref{ga_projector})
and give a definition of the map $P$. The matrix elements
of the map
$P(\nu'',\nu') = \braketfull{\nu''}{\hat{P}}{\nu'}
= \int \! dT \, \braketfull{\nu''}{ e^{-i \hat{H} T} }{\nu'}$ can
be written in terms of a path integral amplitude.
The integrand
looks like an ordinary quantum mechanical propagator and thus
can discretized using
the standard derivation. Decompositions of the identity
(\ref{identity}) are inserted at $n$ discrete time slices
$t_k = k \epsilon$ with time step $\epsilon = \frac{T}{n+1}$ to
get the propagator
\begin{eqnarray}
        \braketfull{\nu''}{ e^{-i \hat{H} T} }{\nu'} &=&
        \lim_{\epsilon \rightarrow 0}
        \prod_{k=1}^{T/\epsilon} \left\{ \sum_{\nu_k, \nu_0} \;
        \int \!\! d\mu_k \;\;
    e^{\,i \sum_{k=1}^{T/\epsilon} \nu_k (A_{k+1}\!-\!A_k) /2 -
        \epsilon H(A_k, \nu_k)}
    \right \} \nonumber \\
    &=& `` \int \mathcal{D}\!A \;\mathcal{D}\!E \; e^{\,i S} \;\;"
\end{eqnarray}
with $\nu_1=\nu'$ and $\nu_{n+1}=\nu''$ and
$S$ being the discretized classical action $S = \sum_{k=1}^{T/\epsilon} \nu_k
(A_{k+1}\!-\!A_k) /2 - \epsilon H(A_k, \nu_k)$.
Notice that the
functional integration over the triad $E$ is discretized as
continuous sums over it's eigenvalues at the time slices $\nu_k/6$
and that $H(A_k, \nu_k) = \frac{24}{\nu_0^2} \sin^2(\nu_0 A_k/2)
\cos^2(\nu_0 A_k/2) +\Lambda \nu_k /2 + \frac{1}{\sqrt{\nu_k/6}}H_{matter}$. We
note that this path integral
amplitude is equivalent to the path integral amplitude arrived
at after gauge fixing due to reparameterization invariance
of the classical action
\footnote{Halliwell \cite{Halliwell:1988wc} has considered the path integral
for actions that are reparameterization invariant such as the action
presented here. Starting with the path integral $P=\int \mathcal{D}A
\, \mathcal{D}E \,\mathcal{D}B \,\exp[i \int^{t''}_{t'} \!\! dt
\,3 E \dot{A}\! +\! B H]$ the
one form component $B$ is gauge fixed to be constant in time and after including
ghost terms to make the path integral independent of the gauge choice the
path integral becomes $P=\int dB (t''\!-\!t') \int \mathcal{D}A
\, \mathcal{D}E \exp[i \int^{t''}_{t'} \! dt \,3 E \dot{A}\! +\! B H]$ which with the
redefinition $T=B (t''-t')$ and $\bar{t}=B(t-t')$
takes on the form equivalent to the group
averaging one $P=\int dT \int \mathcal{D}A
\, \mathcal{D}E \exp[i \int^{T}_{0} \! d\bar{t} \,3 E \dot{A}\! +\! H] =
\int dT \, \braketfull{\nu''}{ e^{-i \hat{H} T} }{\nu'}$
}.

Putting these ideas together, the matrix elements of
the map $P$ are calculated through the path integral as
\begin{equation}
    \label{P}
        P(\nu'',\nu') = \int \! dT \, \lim_{\epsilon \rightarrow 0}
                \prod_{k=1}^{T/\epsilon} \left\{ \sum_{\nu_k, \nu_0} \;
        \int \!\! d\mu_k \;\; e^{\,i S} \right \}  \;.
\end{equation}
We then use this  to map any state $\ket{\psi_0}=\sum_{\nu} {\psi_0}_{\nu}
\ket{\nu} \in \Phi$ onto
a solution of the constraint equation as in
\begin{eqnarray}
    \bra{P(\psi_0)} &=& \int \!dT \bra{\psi_0} e^{-i\hat{H}T}
        \nonumber \\
    &=& \sum_{\nu'', \nu'} \; \braket{\psi_0}{\nu''} \,P(\nu'',\nu') \bra{\nu'}
    \nonumber \\
    &=&  \sum_{\nu'',\nu'} \; \overline{{\psi_0}_{\nu''}}
    \; P(\nu'',\nu') \bra{\nu'}\,.
\end{eqnarray}
The physical inner product between two physical states
$\bra{P(\phi_0)}$ and $\bra{P(\psi_0)}$ is given by
\begin{eqnarray}
    \braket{P(\phi_0)}{P(\psi_0)}_{phys} &=&
        \braketfull{\psi_0}{P}{\phi_0} \nonumber \\
    &=& \sum_{\nu'',\nu'} \; \braket{\psi_0}{\nu''} \;
        \braketfull{\nu''}{P}{\nu'}\; \braket{\nu'}{\phi_0}
        \nonumber \\
    &=& \sum_{\nu'',\nu'} \; \overline{{\psi_0}_{\nu''}} \;
        \; P(\nu'',\nu') \; {\phi_0}_{\nu'} \,.
\end{eqnarray}
With the equivalence relations of equations (\ref{equivstates})
and (\ref{0states}), we have a concrete recipe for characterizing the
physical Hilbert space of the model.

\section{Example - Cosmological Constant}
\subsection{Physical Hilbert Space}
As an example we consider the simplest non-trivial model
with a positive cosmological constant and no other forms
of matter. In this section we will show that this
model can be solved exactly in the connection
representation and we will explicitly construct the
physical Hilbert space. In the following section we
transform to the triad representation in order to discuss the
relationship to loop quantum cosmology.

We have already derived the form of the constraint which
is given by
\begin{eqnarray}
        \hat{H}_{GR}&=& \frac{24}{\nu_0^2} \sin^2(\nu_0 \hat{A}/2)
                \cos^2(\nu_0 \hat{A}/2)
                + 3 \Lambda \hat{E}  \nonumber \\
    &=& \frac{24}{\nu_0^2} \sin^2(\nu_0 \hat{A}/2)
                \cos^2(\nu_0 \hat{A}/2)
                -i \Lambda \frac{\partial}{\partial A} \,.
\end{eqnarray}
Because of the simplicity of the constraint operator we
can solve exactly for the eigenstates which are given
by solutions $\ket{E_n}$ satisfying $\hat{H} \ket{E_n}
= E_n \ket{E_n}$. This differential equation can be solved
to give
\begin{equation}
    E_n(A) = \exp\left[ i \, \frac{ A}{\Lambda}
      \left(E_n-\frac{3}{\nu_0^2}\right) \right]
      \exp \left[ i \, \frac{3 \sin(2\nu_0 A)}{2 \Lambda \nu_0^3}
        \right]
\end{equation}
where the eigenvalues $E_n$ are valued on the real line.  The
eigenstates are pure phase, but they are normalizable in the
kinematical inner product. 
This is despite the fact that any $E_n \in \R$ lies in
the spectrum of $\hat H$.
Since the eigenstates are normalizable
 we expect $P$ to be a bona fide projection operator.  The
kinematical norm is
\begin{eqnarray}
    \braket{E_n}{E_n} &=& \int\!\! d\mu \;\overline{E_n(A)}\;
      E_n(A) \nonumber \\
    &=& \int \!\! d\mu = 1
\end{eqnarray}
and thus all of the eigenstates have kinematical norm equal to one.
Following the general discussion of refined algebraic
quantization, the physical
Hilbert space consists of the space of zero `energy' eigenstates and
the physical inner product is equal to the kinematical one
 restricted to this space.
We see that this space is one dimensional spanned by the
solution $\ket{E_0}$ given by
\begin{equation}
    E_0(A) = \exp \left(
      -i \; \frac{3 A}{ \nu_0^2 \Lambda} +
      i \; \frac{3 \sin(2\nu_0 A)}{2 \nu_0^3 \Lambda} \right) \,.
\end{equation}
We can also write out the matrix elements of the
map $P(A'',A') = \braketfull{A''}{P}{A'}$.
Since the map projects onto the
zero `energy' eigenstate and satisfies $P^2=P$ the
matrix elements must be of the form
\begin{equation}
    P(A'',A') = E_0(A'') \; \overline{E_0(A')}
\end{equation}

We thus see that despite the fact that the kinematical
Hilbert space is very large, the physical Hilbert
space is one-dimensional which is what we expect from
a theory with zero degrees of freedom. In the
next subsection we will see that in the triad
representation there exist a non-countable set
of solutions to the constraint equation---this is in fact generally the case in
standard loop quantum cosmology---but we will 
show that the extra solutions all have zero
physical norm and thus the physical Hilbert
space is one-dimensional.

\subsection{Triad Representation}
In standard loop quantum cosmology a simple formula for
the Hamiltonian constraint in the connection representation
does not exist hence the constraint and solutions are
more readily given in the triad representation. The
space of solutions in standard loop quantum
cosmology is non-separable whereas the physical Hilbert space
of the model presented here is separable consisting of a single
solution. Our aim in this section is to reconstruct the
physical Hilbert space in the triad representation to
facilitate a comparison with standard loop quantum
cosmology. In particular the map
$P$ is calculated explicitly using the path integral
and picks out a single state among all the solutions
to the constraint equation.

In the triad representation the action of the Hamiltonian
constraint operator was given in equation (\ref{constraint_action})
\begin{equation}
    \hat{H} \ket{\nu} = -\frac{3}{2\nu_0^2}
        \Big( \; \ket{\nu \!+\! 4\nu_0 } -2 \ket{\nu}
        + \ket{ \nu \!-\! 4\nu_0 } \;\Big)
    + \frac{\Lambda }{2} \nu \, \ket{\nu} \;.
\end{equation}
We decompose a state lying in the dual of the
kinematical Hilbert space $\bra{\psi}$ as
\begin{equation}
    \bra{\psi} = \sum_{\nu} \psi_{\nu} \bra{\nu} .
\end{equation}
The constraint equation
$ \bra{\psi}H^{\dagger}=0$  leads to a \emph{difference}
equation for the coefficients $\psi_{\nu}$
\begin{equation}\label{diffeqn}
    \psi_{\nu+4\nu_0}-2\psi_{\nu}+\psi_{\nu-4\nu_0} =
    \frac{1}{3} \Lambda \nu_0^2 \; \nu \;\psi_{\nu}.
\end{equation}
It is immediately seen that the difference equation splits
into an infinite number of isolated sectors with values of $\nu$ separated
by an integer times $4\nu_0$. We label each sector by the
parameter $\delta \in [0,4 \nu_0).$ Thus the sector labeled
by $\delta$ corresponds
to values of $\nu$ equal to $\ldots \; \delta,\; 4\nu_0\!+\!\delta,\;
8\nu_0\!+\!\delta \;\ldots$. The difference equation for each sector
is then a second order equation
which can be solved exactly and the general solutions
are given by
\begin{equation}\label{solution}
    \psi_{\nu}^{\delta} = C_1^{\delta}\;  J_{\nu/(4 \nu_0)+
                3/(2 \Lambda \nu_0^3)}
        \left(\frac{3}{2 \Lambda \nu_0^3}\right)
    +C_2^{\delta} \; Y_{\nu/(4 \nu_0)+3/(2 \Lambda \nu_0^3)}
                \left(\frac{3}{2 \Lambda \nu_0^3}\right)
\end{equation}
where $J$ and $Y$ are the Bessel functions of the first and second
kind respectively and $C^{\delta}_{1,2}$ are constants. There are
thus an infinite number of solutions determined by the infinite
number of parameters $C^{\delta}_{1,2}$ for $\delta \in [0,4 \nu_0 )$.

The matrix elements of the map $P$
can be solved exactly for this model using the path integral
(\ref{P}). The discretized action is
\begin{eqnarray}
        S &=& \sum_{k=0}^{T/\epsilon} \frac{\nu_k}{2} (A_{k+1}\!-\!A_k)
        + \epsilon \left[ \, 6 \sin^2(\nu_0 A_k)/\nu_o^2 +
        \Lambda \nu_k /2 \, \right] \nonumber \\
        &=& \sum_{k=0}^{T/\epsilon} \frac{\nu_k}{2} \left[ A_{k+1}\!-\!A_k +
                                \epsilon \Lambda \right] -
                        6 \epsilon \sin^2(\nu_0 A_k) / \nu_0^2
\end{eqnarray}
thus the sums over $\nu_k$ can be performed since they are all of the form
$\exp\left[i\nu_k \left( A_{k+1}\!-\!A_k +
\epsilon \Lambda \right)/2 \right]$ which give delta functions.
The result is
\begin{equation}
        \braketfull{ \nu''}{e^{-i \hat{H} T} }{ \nu'} =
        \lim_{\epsilon \rightarrow 0}
        \prod_{k=1}^{T/\epsilon} \Bigg\{
        \begin{array}{ll}
        \int^{\infty}_{-\infty}  dA_k &
            \delta(A_{k+1}\!-\!A_k+\epsilon \Lambda) \;
            e^{i\nu''A_n/2}\,e^{-i\nu' A_0/2}  \\
        &e^{i\sum_{k=0}^n 6\epsilon \sin^2 (\nu_0 A_k) / \nu_0^2}
    \end{array}
        \Bigg\} \,.
\end{equation}
Let us then integrate over the $A_k$ using the delta functions
except for $A_0$ and $A_n$.
This leaves us with one remaining delta function
\begin{equation}
        \braketfull{\nu''}{e^{-i \hat{H} T}}{ \nu'} =
    \begin{array}{ll}
        \int^{\infty}_{-\infty} \! dA_0 \,dA_n &
            \delta(A_n\!-\!A_0+T \Lambda) \,e^{i\nu''A_n/2}\,
        e^{-i\nu' A_0/2}  \\
            & \exp \left[ \frac{3 i}{ \nu_0^2} \left( T - \frac{\sin(T \nu_0 \Lambda)
                \cos(2\nu_0 A'-T \nu_0 \Lambda)}{\nu_0 \Lambda} \right)
            \right]
    \end{array} \;\;.
\end{equation}
The path integral is then given by
\begin{eqnarray}\label{projector}
        P(\nu'',\nu') &=& \int^{\infty}_{-\infty} \! dA_0 \,dA_n
    \,e^{i\nu''A_n/2}\,  e^{-i\nu' A_0/2}
    \int \!\! dT  \; \delta(A''\!-\!A'+T \Lambda) \;
        e^{\frac{3i}{\nu_0^2} \left[ T - \frac{\sin(T \nu_0 \Lambda)
                \cos(2\nu_0 A'-T \nu_0 \Lambda)}{\nu_0 \Lambda} \right]}
        \nonumber \\
        &=&  \int^{\infty}_{-\infty} \! dA_0 \,dA_n
    \,e^{i\nu''A_n/2}\,  e^{-i\nu' A_0/2} \,
    e^{ \left( \frac{3 i\sin(2\nu_0 A'')}{2 \nu_0^3 \Lambda} -
        \frac{3i A''}{ \nu_0^2 \Lambda} \right)} \;
        e^{ \left(-\frac{3i\sin(2 \nu_0 A')}{2 \nu_0^3 \Lambda} +
        \frac{3iA'}{ \nu_0^2 \Lambda} \right) }
        \nonumber \\
    &=&  \int^{\infty}_{-\infty} \! dA_0 \,dA_n
    \,e^{i\nu''A_n/2}\,  e^{-i\nu' A_0/2}
    \, E_0(A'') \; \overline{E_0(A')}
\end{eqnarray}
which is just the Fourier transform of the map from the previous section.
$P$ thus maps onto the single normalizable zero 'energy' eigenstate
$\ket{E_0}$.

We now formulate the physical Hilbert space by using the
map (\ref{projector}). $P$ maps onto a single state in
the triad representation which is given by the Fourier
transform of the zero 'energy' eigenstate $\ket{E_0}$.
Denoting $\ket{E_0} = \sum_{\nu} p_{\nu}
\ket{\nu}$ a calculation gives for the Fourier transform
\begin{equation}\label{normedsol}
        p_{\nu}= \left\{
        \begin{array}{ll}
                J_{\nu/(4 \nu_0)+3/(2 \Lambda \nu_0^3)}
                \left(\frac{3}{2 \Lambda \nu_0^3}\right)
                \;\;\;\;\;\;&
                \frac{\nu}{4 \nu_0}+\frac{3}{2 \Lambda \nu_0^3}
                \in \mathbb{Z} \\
                0 &   \frac{\nu}{4 \nu_0}+\frac{3}{2 \Lambda \nu_0^3}
                \notin \mathbb{Z} .
        \end{array}
        \right.
\end{equation}
Remarkably, the path integral solution is non zero in precisely
\emph{one} sector $\delta_c$ corresponding to values of $\nu$ where the
Bessel function order $\frac{\nu}{4 \nu_0}+ \frac{3}{2 \Lambda
\nu_0^3}$ is an integer. The value of the special sector is
thus given by
$\delta_c=4 \nu_0\, [1 -  mod(\frac{3}{2 \Lambda\nu_0^3}, 1)]$
where $mod(a,b)$ is the remainder of $a/b$. 
That the Fourier transform picks out certain modes
 can be seen from the fact that the state
$E_0(A)$ consists of a phase factor multiplied by a periodic
function. The matrix elements of $P$ in the triad representation
are thus given by $P_{\nu \nu'} = \braketfull{\nu}{P}{\nu'} =
p_{\nu} \, \overline{p_{\nu'}}$. Using this form of the map, any
kinematical state will get projected onto the zero 'energy'
eigenstate given in equation (\ref{normedsol}).
The uncountable extra solutions (\ref{solution}) to the
constraint equation in the triad representation are to be mod out
of the physical Hilbert space. In this way the projector has
reduced what was initially a non-separable space of solutions to
the constraint equation to a one dimensional Hilbert space. This is
because the generic solutions of the constraint correspond to {\em
zero} physical norm states and therefore are equivalent to the
{\em zero} state in the physical Hilbert space.

The previous statement requires a more precise explanation as the
generic solutions (\ref{solution}) are not normalizable in terms
of the measure (\ref{measure}). These solutions have {\em zero
physical norm} in the following sense: The generic solutions
(\ref{solution}) can be thought of as elements of $\Phi^{\star}$
defined in Section \ref{rac}.  Given a generic solution
$\psi^{\delta}\in \Phi^{\star}$ (as in (\ref{solution})) 
with values only in the sector $\delta$, we define
the shadow state $\psi^{\delta}_{N}$ with $N \in \N$ as
\begin{equation}
        \psi^{\delta}_{N}= \left\{
        \begin{array}{ll}
                 C_1^{\delta}\;  J_{\frac{\nu}{4 \nu_0}+
                \frac{3}{2 \Lambda \nu_0^3}}
        	\left(\frac{3}{2 \Lambda \nu_0^3}\right)
    	+C_2^{\delta} \; Y_{\frac{\nu}{4 \nu_0}+\frac{3}{2 \Lambda \nu_0^3}}
                \left(\frac{3}{2 \Lambda \nu_0^3}\right) \;\;
		& {\rm for}\,\, \nu=4n\nu_0+\delta\,\, {\rm with}\,\, n\in[-N,N]\\
                0 &   {\rm elsewhere}.
        \end{array}
        \right.
\end{equation}
Clearly the shadow state $\psi^{\delta}_{N}$ is an 
almost periodic function and therefore $\psi^{\delta}_{N}\in
\Phi$. Notice also that as $N\rightarrow \infty$, 
$\psi^{\delta}_{N}$ approaches $\psi^{\delta}\in \Phi^{\star}$.
The statement that the generic solutions have zero norm
corresponds to the statement that
\begin{equation}\label{zero}
 \braket{P  (\psi^{\delta}_N)}{P (\psi^{\delta}_N)}_{phys}=
 0  
\end{equation}
for  $\delta \neq \delta_c$  
where again $\delta_c=4 \nu_0\, [1 -  mod(\frac{3}{2 \Lambda\nu_0^3}, 1)]$ 
according to (\ref{normedsol}).

From the form of equation (\ref{projector}) it is clear that
$P^2=P$ and hence the kinematical inner product can be used as the
physical inner product. We can see explicitly that the solution
(\ref{normedsol}) is normalizable by calculating it's norm
\begin{eqnarray}
    \braket{E_0}{E_0} &=& \sum_{\nu} \; \overline{p_{\nu}}\; p_{\nu}
        \nonumber \\
    &=& \sum^{\infty}_{m=-\infty}
        \left[ J_m\left({\scriptstyle \frac{3}{2 \Lambda \nu_0^3}}
        \right) \right]^2 = 1
\end{eqnarray}
using a known formula for the sum over Bessel functions of integer
order.

\begin{figure}[ht]
\begin{center}
  \includegraphics[width=9cm,height=6cm,keepaspectratio]
        {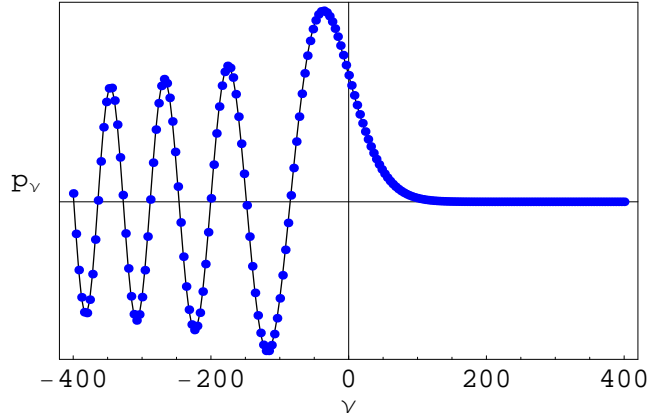}
\end{center}
\caption{Physical solution (\ref{normedsol}). The dots indicate
the values of the solution for where $\frac{\nu}{4 \nu_0}+
\frac{3}{2 \Lambda \nu_0^3}\in \mathbb{Z} $. The actual
solution would be zero elsewhere, however the solid lines are
interpolated values designed to bring out the behavior of the solution
more clearly.}
\label{Euclidean_smallvu}
\end{figure}

The physical solution (\ref{normedsol}) is plotted in figure
(\ref{Euclidean_smallvu}). We see that despite the discrete nature
of the solution, it approximates a continuous one which for large
values of $\nu$ would approximate the Wheeler-DeWitt equation. For
positive values of $\nu$ the solutions decays rapidly which would
be expected for the Riemannian model which has no classically
allowed region for an effective 
\footnote{Notice that the effect of the cosmological constant
term depends on the orientation of the triad which is dynamical
for our action.} positive cosmological constant. 
For negative $\nu$ the solution oscillates
with behavior indicating an effective negative cosmological constant. 
In the discussion that
follows we will indicate why this occurs and the interpretation
for the Lorentzian regime. Furthermore, we can see that the
classical singularity ($\nu = 0$) does not present a barrier to
the evolution as the wave function can be evolved through
it.

\section{Discussion}
The simplicity of the Hamiltonian constraint used here has allowed
for explicit calculations to be performed. A rigorous definition of the
physical scalar product was given and used to factor out spurious 
solutions 
to the constraint equations with zero physical norm. 
We wish to compare the results of
the model with those of standard loop quantum cosmology. To do
so, we observe that there is a simple relationship between
the Riemannian and Lorentzian theories for homogeneous
and isotropic models. Using this simple relation we can compare the two
Hamiltonian constraints and the physical implications of the differences.
In addition, we will discuss the cosmological predictions of such
a model.

\subsubsection*{The Lorentzian model and relation
to standard LQC}
A Lorentzian model can be constructed if one uses general
properties of the isotropic reduction of general relativity. 
With homogeneity and isotropy the curvature 
part of the Hamiltonian constraint changes sign while the cosmological
constant remains the same when going from the Riemannian sector
to the Lorentzian one. Using this property the
Hamiltonian constraint for the Lorentzian constraint becomes
$H_L = -6 A^2 + 3 \Lambda E$ as
opposed to the Riemannian constraint $H_E = 6 A^2 + 3 \Lambda E$.
Equivalently the same effect is obtained by 
changing the sign of the triad $E$ (which amounts to sending
$\nu$ to $-\nu$ in the solutions presented here). 

If we extend our model in such a fashion the difference equation
becomes
\begin{equation}\label{diff_our}
    \psi_{\nu+4\nu_0}-2\psi_{\nu}+\psi_{\nu-4\nu_0} =
        - \frac{1}{9}  \gamma^3 \nu_o^2 l_p^2
    \, \Lambda \; \nu \;\psi_{\nu}
\end{equation}
where we have included the various constants needed to compare to
standard loop quantum cosmology. Here
$\gamma$ is the Barbero-Immirzi parameter, $l_p = \sqrt{\kappa \hbar}$ 
is the Planck length, and $\kappa = 8\pi G$.
The difference equation for standard loop quantum cosmology is
given by
\begin{eqnarray}\label{diff_LQC}
    \left( V_{\nu+5\nu_0}\!-\!V_{\nu+3\nu_0} \right) \psi_{\nu+4\nu_0}\,-\,
    2 \left( V_{\nu-\nu_0}\!-\!V_{\nu+\nu_0} \right) \psi_{\nu} \,+\,
    \left( V_{\nu-3\nu_0}\!-\!V_{\nu-5\nu_0} \right) \psi_{\nu-4\nu_0}
    \nonumber \\
    = -\frac{1}{3} \gamma^3 \nu_0^3 l_{p}^2\, \Lambda
    \,sgn(\nu)\, V_{\nu} \, \psi_{\nu}
\end{eqnarray}
where the volume eigenvalues are $V_{\nu}=(|\nu| \gamma l_p/6)^{3/2}$,
and $sgn$ stands for the sign function. That the two constraints
are asymptotically the same  can be shown
by using the large volume expansion for the volume difference coefficients
$V_{\nu + \nu_0}\!-\!V_{\nu-\nu_0} \approx
\left( \gamma l_p / 6  \right)^{3/2} 3 \sqrt{\nu} \nu_0$. Plugging
this approximation into (\ref{diff_LQC}) then gives the difference
equation (\ref{diff_our}). The two models would thus share the same
large volume semi-classical limit.

The differences between the two difference equations can be traced back
to the classical Hamiltonian constraints used. The constraint
of standard loop quantum cosmology is
\begin{equation}
    H_{LQC} = -\frac{6}{\kappa \gamma^2} A^2 sgn(E) \sqrt{|E|}+
         \frac{\Lambda}{\kappa} sgn(E)(|E|)^{3/2}
\end{equation}
whereas the constraint used here would be
\begin{equation}\label{Hclass}
    H = -\frac{6}{\kappa \gamma^2} A^2 +  \frac{\Lambda}{\kappa} E \,.
\end{equation}
It is readily seen that for positive values of the triad the two constraints
are proportional to
each other and thus are classically equivalent on the constraint surface where
$H=0$. The main qualitative difference between the two constraints is
that they are not equivalent for negative orientations of the
triad. This is indicated in the difference equation by the presence of
the $sgn(\nu)$ term on the right hand side of equation (\ref{diff_LQC}).
The factor $sgn(\nu)$ is put in by hand in standard LQC such that the constraint is classically
symmetric for both orientations of the triad. However, there is a
freedom in extending the classical phase space to include negative
orientations of the triad. We have shown that 
the action (\ref{full_BF_action}) does not lead to the $sgn(\nu)$ term.
In the next subsection we discuss in more detail the 
implications of such a change.

The other crucial difference between the constraints is
that our constraint is self-adjoint. This allows us to carry out
the group averaging technique to construct the physical
inner product. The constraint of standard
loop quantum cosmology is not, although self-adjoint constraints
have been proposed \cite{Bojowald:2004zf, Willis:thesis}.

\subsubsection*{The triad orientation ambiguity}
We have stated the main qualitative difference between
the simpler model lies in the classical extension
to negative orientations of the triad. This
ambiguity, which is manifested by the presence of the $sgn(\nu)$ term
in the difference equation (\ref{diff_LQC}), has important
consequences in the quantum theory. We now show
that absence of the $sgn(\nu)$ term --- forced on
us by the action (\ref{full_BF_action}) ---
leads to many attractive features in the quantum
theory.

\begin{figure}[ht]
\centering
\mbox{
    \subfigure[]
        {\epsfig{figure=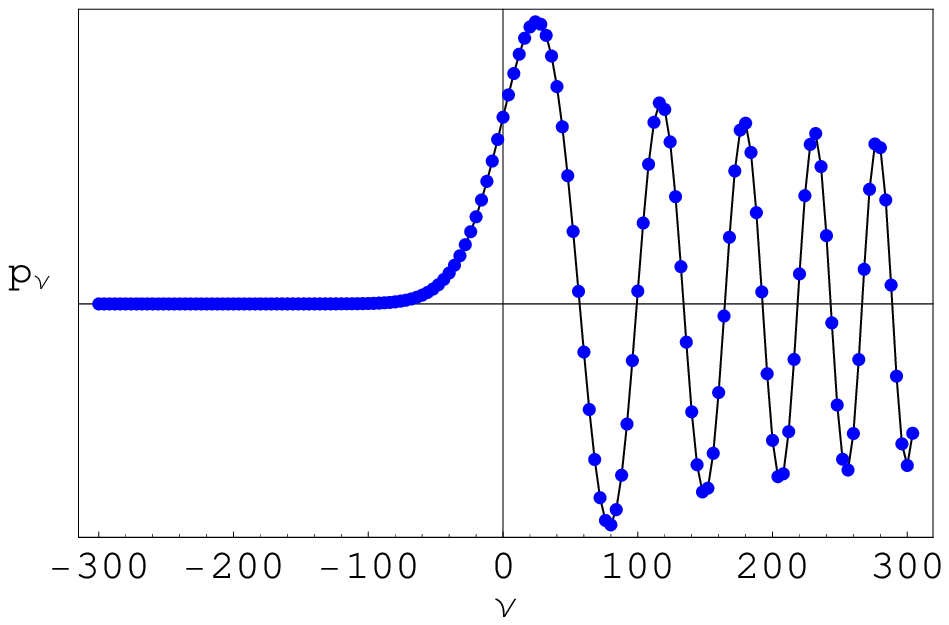,width=8cm,keepaspectratio}
        \label{L_smalla}}
    \subfigure[]
        {\epsfig{figure=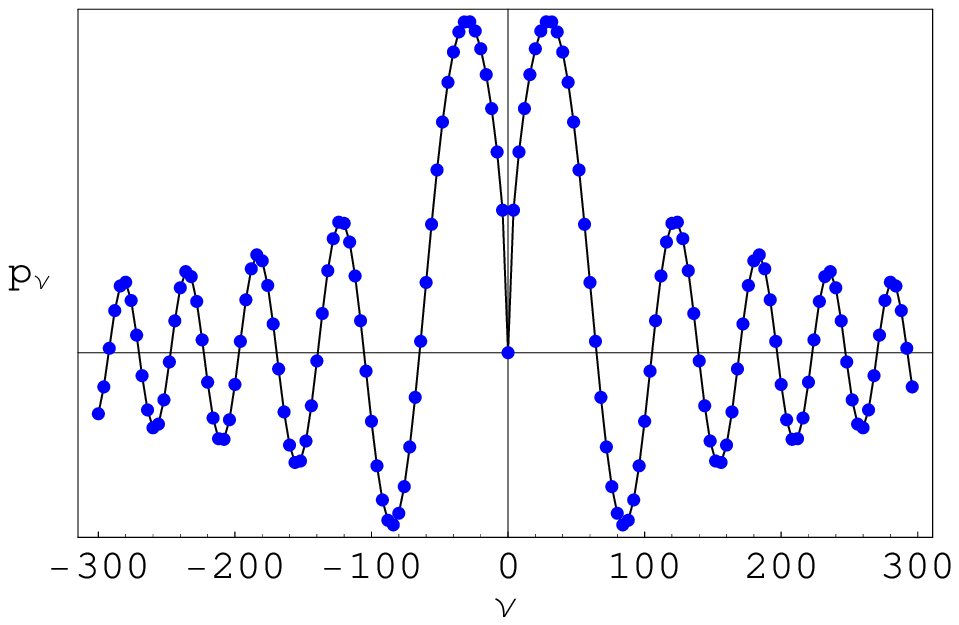,width=8cm,keepaspectratio}
        \label{L_smallb}}
}
\caption{Difference equation solutions for both the simplified constraint
(a)
and standard loop quantum cosmology (b). The solution of (a) is the
physical solution. The solution in (b) is the one picked out
by the dynamical initial condition. The solid line are the interpolated
values. The actual physical solution of (a) would be zero
except for the one sector indicated by the dots. For (b)
the dynamical initial condition specifies that the wave function
in the other sectors is the one that smoothly interpolates between the
dotted solution.}
\label{L_small}
\end{figure}

The physical wave function solutions of the two difference equations
are shown in figure \ref{L_small}. For large positive values of the
triad (positive $\nu$) the solutions behave the same.  However, for
negative $\nu$ the solutions are entirely different owing to the
$sgn(\nu)$ term.
Because of that term, the solution in figure \ref{L_smallb}
is symmetric and thus both orientations of the triad have the same
semi-classical limit.  On the other hand, the solution of the
simplified constraint clearly does not have the right semi-classical
limit for negative orientations of the triad. At first one
might think that this is undesirable in the quantum theory since a
large portion of the quantum configuration space does not have the
right semi-classical limit. However, we see that the solution for
negative triad is rapidly suppressed indicating a classically
forbidden region. Thus, the universe would have very low probability
for being in the region and would instead more likely be found in the
semi-classical regime of positive $\nu$.  Notice that because $P^2=P$,
the standard notion of probability amplitude can be associated to the
physical wave function for the single partial observable $E$.
Physically the cosmological constant term in (\ref{Hclass}) changes its sign
when $E$ goes to $-E$. The regions $\nu > 0$ and $\nu<0$ 
correspond to an effective positive and negative cosmological
constant respectively.

\begin{figure}[ht]
\centering
\mbox{
        \subfigure[$-\nu/(4 \nu_0)+ 9/(2 \gamma^3 l_p^2 \Lambda \nu_0^3)
\in \mathbb{Z}$]
                {\epsfig{figure=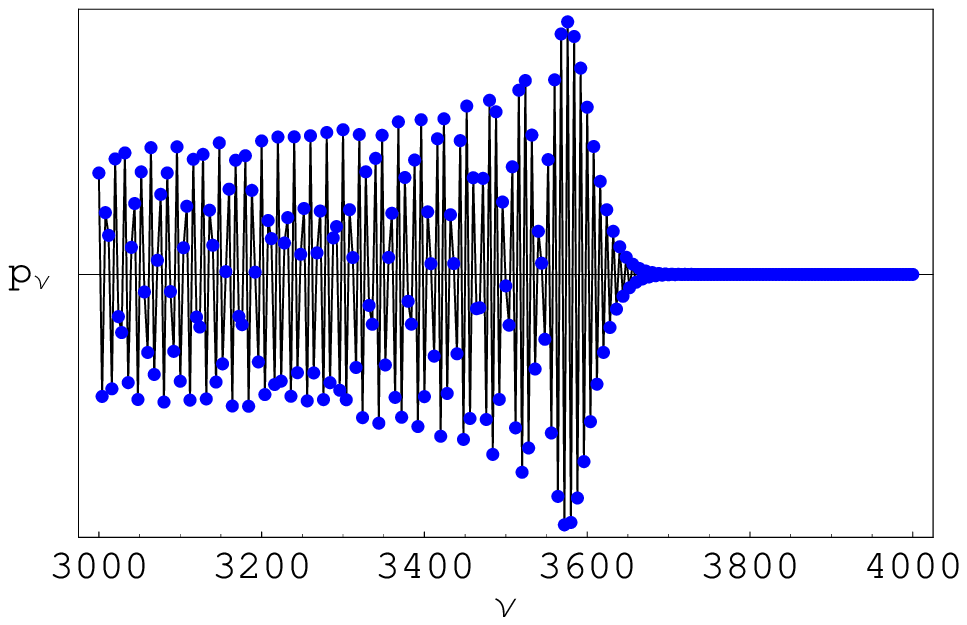,width=8cm,keepaspectratio}
    \label{L_largea}}
        \subfigure[$-\nu/(4 \nu_0)+ 9/(2 \gamma^3 l_p^2 \Lambda \nu_0^3)
\not\in \mathbb{Z}$]
                {\epsfig{figure=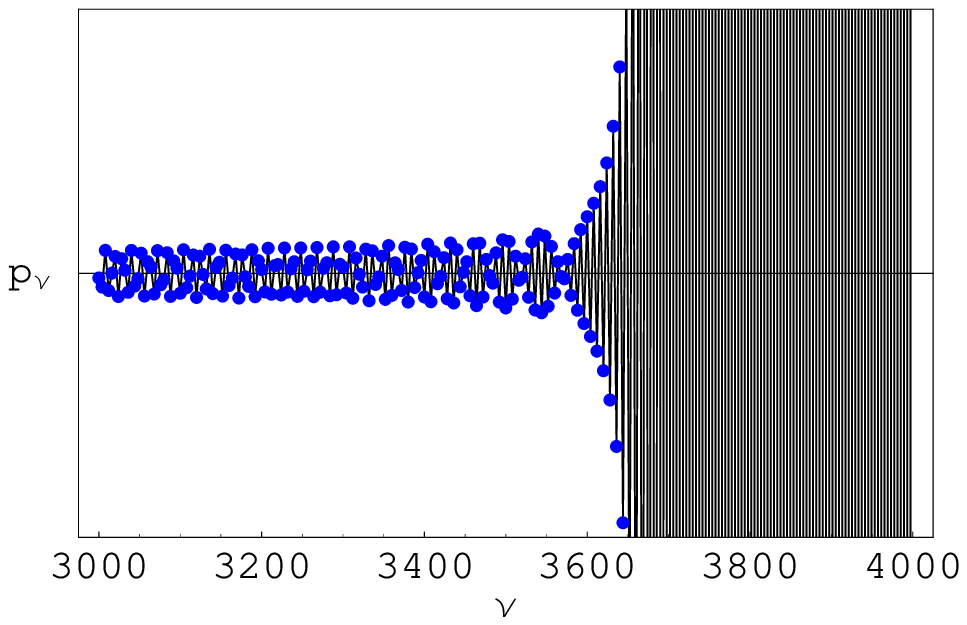,width=8cm,keepaspectratio}}

}
\caption{Bessel J solution of (\ref{Lsolution}) for two different
sectors. The solution of (a) is for the sector where the Bessel function
order is an integer. This solution converges both for large negative and
positive $\nu$. The solution of (b) is for a different sector and diverges
for large $\nu$. The solid lines are interpolated values.
}
\label{normalizability}
\end{figure}

It is precisely because of the absence of the $sgn(\nu)$ term 
that the model can be solved exactly. If we were to use a 
constraint with the $sgn(\nu)$ term the path integral
would not be solvable explicitly and it can
be shown that there do not exist any kinematically
normalizable solutions. The fact that the physical
solution is kinematically normalizable simplified
the construction of $\mathcal{H}_{phys}$.
Without kinematically normalizable solutions
the path integral needs to be solved in order to
construct the physical inner product. The normalizability of
the solutions can be
better understand by looking at the
large volume behavior of the solutions. As in the Riemannian model
the general solution of the simplified model is
\begin{eqnarray}\label{Lsolution}
        \psi_{\nu}^{\delta} = &&C_1^{\delta}\;  J_{-\nu/(4 \nu_0)+
                                9/(2 \gamma^3 l_p^2 \Lambda \nu_0^3)}
                \left(\frac{9}{2  \gamma^3 l_p^2 \Lambda \nu_0^3}\right)
        \nonumber \\
    &+&C_2^{\delta} \; Y_{-\nu/(4 \nu_0)+
                9/(2 \gamma^3 l_p^2 \Lambda \nu_0^3)}
                \left(\frac{9}{2  \gamma^3 l_p^2 \Lambda \nu_0^3}\right)
    \,.
\end{eqnarray}
The Bessel Y solutions all diverge for negative $\nu$ and the
Bessel J solutions all diverge for positive $\nu$ except for the
one sector where the Bessel function order is an integer as it was
in the Riemannian solution (\ref{normedsol}). The Bessel J
solutions are plotted for large $\nu$ in figure
\ref{normalizability}. It is clear that only in the special sector
$\delta = \delta_c$ does the solution not diverge. 
If we were to include the $sgn(\nu)$
term then the oscillatory solutions would exist for positive and
negative $\nu$, however we could not match a convergent solution
for positive $\nu$ with one for negative $\nu$. We thus see that
without the $sgn(\nu)$ term, the model has very attractive
features and the construction of the physical Hilbert space is
greatly simplified since there exists a normalizable solution.

A question remains as to the relevance of the simplified constraint to
that of standard LQC: are these special results specific to the
simplified model? If we remove the $sgn(\nu)$ term from
the difference equation of \emph{standard LQC}, it can be shown
that the model with  a cosmological constant does
indeed exhibit similar behavior. To show this, the
difference equation must be solved numerically since it is more
complicated.  Consider that we start at large negative $\nu$ with some
initial conditions of the wave function and we evolve forward toward
the classical singularity. For large negative $\nu$ the two
independent solutions of the difference equation behave in a
characteristic manner.  The difference equation for large $|\nu|$ is
approximated by the difference equation of our simplified model.
Therefore, for large and negative $\nu$ the two independent solutions
behave as $f_+(\nu)\approx \nu^{-(\nu+1/2)}$ and $f_-(\nu)\approx
\nu^{\nu-1/2}$ which can be shown from the asymptotic expressions of
the solutions in equation (\ref{Lsolution}).  For generic initial
conditions the solution will be a linear combination of $f_+$ and
$f_-$. As we evolve toward the singularity $f_+$ rapidly dominates
over $f_-$. In this manner numerically the contribution of the
decaying solution $f_-$ becomes negligible in comparison with the
contribution of $f_+$ thus effectively selecting the latter.
We can perform the same trick starting at large
positive $\nu$---where the independent solutions behave as
$g_+(\nu)\approx (-1)^{\frac{\nu}{4\nu_0}}\, \nu^{(\nu-1/2)}$ and
$g_-(\nu)\approx (-1)^{\frac{\nu}{4\nu_0}}\, \nu^{-(\nu+1/2)}$---and evolving backward to select the $g_+$ component.
The two solutions can then be tested to see
if they match somewhere in the region where the behavior is
oscillatory. If the two can be  matched then we have found
a solution that decays both for positive and negative
$\nu$.

\begin{figure}[ht]
\begin{center}
  \includegraphics[width=9cm,height=6cm,keepaspectratio]
        {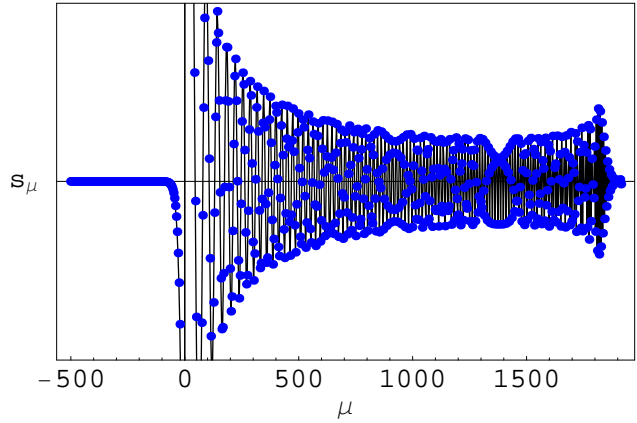}
\end{center}
\caption{Normalizable solution to the difference equation
of LQC (\ref{diff_LQC}) without the $sgn(\nu)$ on the
r.h.s. Here matter is only in the form of
a cosmological constant. The solution is valued
only in a single sector which approximately is
given by the same sector of the simplified constraint
where $-\nu/(4 \nu_0)+ 9/(2 \gamma^3 l_p^2 \Lambda \nu_0^3)
\in \mathbb{Z}$.
}
\label{LQC_norm}
\end{figure}

When this analysis is performed for the difference
equation of LQC with a cosmological constant (without the $sgn(\nu)$ term)
the result is a match
only in one sector which corresponds approximately to
the one picked out by our simplified model and given
by $-\nu/(4 \nu_0)+ 9/(2 \gamma^3 l_p^2 \Lambda \nu_0^3)
\in \mathbb{Z}$. In the other sectors the matching cannot be achieved
which implies for instance that a solution that decays toward $-\infty$ will evolve
into one that diverges toward $+\infty$. Since the falloff behavior at
large positive and negative $\nu$ is sufficiently fast we can say that
the solution that does not diverge is kinematically normalizable 
as in our simplified model.  The
solution is plotted  in figure
(\ref{LQC_norm}). It has the same qualitative behavior as the
normalizable solution of the simplified constraint (compare to figures
\ref{L_smalla} and \ref{L_largea}).

We are currently investigating whether a normalizable solution exists
for a massless scalar field in the closed model.  A recent paper has
pointed out that in this model all the solutions diverge either for
large positive or negative $\nu$ \cite{Green:2004mi}.  The 
question is raised that this implies that the wave function
then predicts a large `probability' for the universe being in those
regions which are classically forbidden for this model.  Notice however
that any question about probabilities can only be addressed if a
notion of physical inner product is provided. The physical inner
product can solve this apparent problem by providing a non trivial
probability measure that makes the physical states normalizable.  We
have shown for our model that most solutions have divergent behavior
yet the physical inner product singles out a solution---
all the other ill-behaved solutions have zero norm---which is
finite everywhere and even kinematically normalizable. Therefore it is
interesting to explore whether this might occur with the model of
\cite{Green:2004mi}.  If a normalizable solution can be found
which decays for large volume then the problems raised would be shown
to be nonexistent. Preliminary results indicate that one might be able to find
normalizable solutions for this model again provided the $sgn(\nu)$
term is removed.

\subsubsection*{Cosmological implications}
We now discuss the cosmological implications of the Lorentzian
model. A main question is whether or not the quantum theory
cures the classical singularity. A classical singularity
would exhibit itself in the quantum theory in two ways: a
place where the difference equation breaks down or a place
where the operator corresponding to the inverse scale factor is
singular. For both these criteria the model presented here does
not exhibit singular behavior. We see from the figure \ref{L_smalla}
that the quantum evolution proceeds smoothly through the classical
singularity ($\nu = 0$). From the standpoint of the difference equation
the singularity point is not special. In addition the operator corresponding
to the inverse scale factor is a kinematical operator which has been shown
to be bounded in standard loop quantum cosmology \cite{Bojowald:2001vw}.
Since the model
here shares the same kinematical Hilbert space, the results will
be the same.
We furthermore see that the physical solution is valued only
on discrete values of $\nu$ corresponding to the special sector
$\delta_c$.
The special sector depends on the value of the cosmological constant
and only specific values will select the sector $\delta=0$ which
passes through
$\nu=0$. Thus, it is most likely that the physical solution
completely avoids the singularity. The singularity is thus cured
in the model.

If the model does not exhibit singular behavior then another
question is what happens when the universe approaches the
singularity. The plot in figure \ref{L_smalla} shows that the
solution decays rapidly for negative values of $\nu$ while
it oscillates for positive values. Thus it is natural to interpret
the region of negative $\nu$ as a classically forbidden region and
the singularity as a barrier off which the universe bounces. This is
in contrast to standard loop quantum cosmology where semi-classical
universes exist for both negative and positive $\nu$ thus opening
the possibility for the universe to tunnel through the singularity
from one region to the other. Again, the main difference arises from
the $sgn(\nu)$ term in the difference equation. From this
we can make the interpretation that a classical collapsing
universe approaching the singularity completely avoids it
and bounces leading to an expanding universe.

Finally, there is the issue of boundary proposals. In
Wheeler-DeWitt quantization there exist two independent solutions
for de Sitter space 
and ad-hoc boundary proposals are added to pick out a solution.
In standard loop quantum cosmology the dynamical initial condition
is used to pick out a solution. However, the dynamical initial condition
only singles out a solution for the sector that passes through the
singularity. Semi-classical arguments are then used to fix the
wave function in the other sectors (see figure \ref{L_smallb}).
In contrast we see that for the simplified model here, it is the
physical inner product that selects a unique wave function.
There is no need to supplement the quantum theory with an ad-hoc
boundary proposal. This is not surprising for this
model as the constraint equation is first order in
the connection representation. The question remains
as to whether this is relevant in more general
situations.

\subsubsection*{Further implications for LQC}

At this stage and in the context of the model presented here one
question seems natural: what have we gained by setting up the theory
on the non separable kinematical Hilbert space given by the Bohr
compactification of the real line? The answer is clear in that
 considering quasi-periodic functions of $A$ is the analog of
considering cylindrical functions of the connection defined on
arbitrary graphs in the full theory. Had we defined the quantum theory
by using functions of a fixed periodicity (allowing $\nu=\nu_0 n$ for
$n\in \Z$) we would have missed the physical solution $E_0(A)$ and the
physical Hilbert space would have been $0$-dimensional. Only in the
special case when $(3/2)\Lambda^{-1} \nu_0^{-3}$ is an integer could
we have found the physical state by starting with a formulation on a
fixed `lattice', i.e., where $\nu=\nu_0 n$ (see Equation
(\ref{normedsol})). In any other case using a fixed lattice would have
resulted in a zero dimensional physical Hilbert space. In our model,
the physical inner product selects a given set of `graphs' by
selecting a periodicity of the relevant modes. Thus the
Bohr compactification is necessary to capture the correct physics.

Another related issue is that the physical Hilbert space
separates into orthogonal subspaces.
In the standard quantizations of the
Hamiltonian constraint in the literature the curvature
term is written in terms of holonomies along
paths whose length is determined by the parameter
$\nu_0$. Due to this fact one introduces an
intrinsic periodicity to the constraint.
It is easy to see
that the generalized projection operator associated to 
such a quantization satisfies the following property
\begin{equation}
       \braketfull{ \nu''}{P }{ \nu'}= \int dT\ \braketfull{ \nu''}{e^{-i \hat{H} T} }{ \nu'}=0
\end{equation}
for $\nu''-\nu' \notin 4\nu_0 n$ for $n\in\Z$.
In this manner the physical Hilbert
space is separated into isolated sectors with no
quantum interference between them.

\begin{figure}[ht]
\begin{center}
  \includegraphics[width=9cm,height=6cm,keepaspectratio]
        {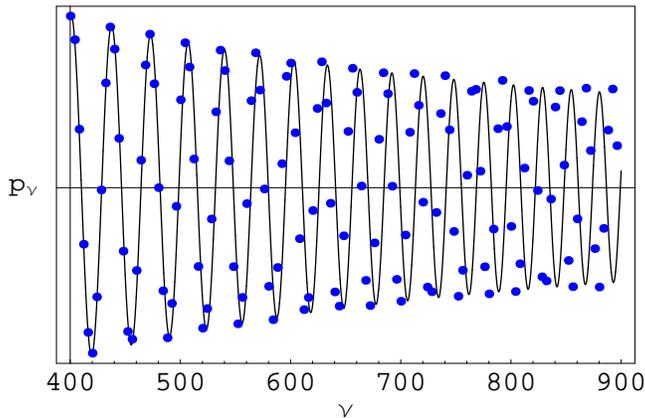}
\end{center}
\caption{Wave function of standard LQC (dots) compared
with the corresponding Wheeler-DeWitt solution (solid line)
for positive cosmological constant. The deviation occurs
when the extrinsic curvature is of the order of
$\pi/\nu_0$ which in this case occurs for large
volumes.
}
\label{LQC_WD}
\end{figure}

A direct consequence of this is the fact that 
in a given sector the physical solutions have a
built in periodicity in that $\Delta \nu=4\nu_0$
which implies that physically $A\le A_{max}= \pi/ \nu_0$. 
This is a puzzling feature as $A$ is
directly related to the extrinsic curvature which classically does
not have such a bound. In
the model with a cosmological constant, classically
the extrinsic curvature grows as the volume of the universe increases
until at some volume it reaches a critical value corresponding to
$A =  A_{max}= \pi/ \nu_0$. At  values of $\nu$ on 
the order of this critical value the behavior of the wave
function changes dramatically and deviates
from that expected from standard Wheeler-DeWitt quantization.
This is a generic property that applies to all models
of LQC (see figure \ref{LQC_WD}).
However, it is in this region that both prescriptions should
coincide.  By decreasing the value of $\nu_0$
one can extend the region where LQC and Wheeler-DeWitt quantization
agree, yet this parameter is argued to be fixed by the full theory
by considering the smallest area eigenvalue \cite{Ashtekar:2003hd}. It is
surprising that a parameter arising from the fundamental
discreetness at Planck scale should have such a important
effect at large scales where the universe is expected to behave
classically.

\section*{Acknowledgments}
We would like to thank Martin Bojowald for proposing this problem and
Abhay Ashtekar and Martin Bojowald for useful discussions. 
This work has been supported by NSF grants PHY-0354932 and INT-0307569 and the
Eberly Research Funds of Penn State University.


\end{document}